\documentclass[10pt,conference]{IEEEtran}
\IEEEoverridecommandlockouts
\usepackage[noadjust]{cite}
\usepackage{amsmath,amssymb,amsfonts}
\usepackage{algorithmic}
\usepackage{graphicx}
\usepackage{textcomp}
\usepackage{xcolor}
\usepackage{colortbl}
\usepackage{tabularx, makecell, multirow, balance, booktabs}
\usepackage{url}
\usepackage{array}
\usepackage{hhline}
\usepackage[hidelinks]{hyperref}
\usepackage[numbers]{natbib} %For the citet

\usepackage{highlight}%From an independent dev.
\renewcommand{\opacity}{65}

\newcolumntype{C}[1]{>{\centering}m{#1}}

\usepackage{tikz}
\usepackage{lipsum}

\newlength{\RoundedBoxWidth}
\newsavebox{\GrayRoundedBox}
\newenvironment{GrayBox}[1][\dimexpr\linewidth-4.5ex]%
   {\setlength{\RoundedBoxWidth}{\dimexpr#1}
    \begin{lrbox}{\GrayRoundedBox}
       \begin{minipage}{\RoundedBoxWidth}}%
   {   \end{minipage}
    \end{lrbox}
    \begin{center}
    \begin{tikzpicture}%
       \draw node[draw=black,fill=black!10,rounded corners,%
             inner sep=2ex,text width=\RoundedBoxWidth]%
             {\usebox{\GrayRoundedBox}};
    \end{tikzpicture}
    \end{center}}

\def\BibTeX{{\rm B\kern-.05em{\sc i\kern-.025em b}\kern-.08em
    T\kern-.1667em\lower.7ex\hbox{E}\kern-.125emX}}
\begin{document}

\title{Developers Need Protection, Too: \\Perspectives and Research Challenges for Privacy in Social Coding Platforms}

% \author{\IEEEauthorblockN{*Anonymous Authors*}
% \IEEEauthorblockA{
% \\
% \\
% \\
% \\
% \\
% }}

\author{\IEEEauthorblockN{Nicolás E. Díaz Ferreyra}
\IEEEauthorblockA{\textit{Institute of Software Security} \\
\textit{TU Hamburg}\\
Hamburg, Germany \\
nicolas.diaz-ferreyra@tuhh.de}
\and

\IEEEauthorblockN{Abdessamad Imine}
\IEEEauthorblockA{\textit{Lorraine University and Loria-} \\
\textit{Cnrs-Inria Nancy-Grand Est}\\
Nancy, France \\
abdessamad.imine@loria.fr}
\and

\IEEEauthorblockN{Melina Vidoni}
\IEEEauthorblockA{\textit{School of Computing} \\
\textit{Australian National University}\\
Canberra, Australia \\
melina.vidoni@anu.edu.au}
\and

\IEEEauthorblockN{Riccardo Scandariato}
\IEEEauthorblockA{\textit{Institute of Software Security} \\
\textit{TU Hamburg}\\
Hamburg, Germany \\
riccardo.scandariato@tuhh.de}}

%\raggedbottom

\maketitle

\begin{abstract}
Social Coding Platforms (SCPs) like GitHub have become central to modern software engineering thanks to their collaborative and version-control features. Like in mainstream Online Social Networks (OSNs) such as Facebook, users of SCPs are subjected to privacy attacks and threats given the high amounts of personal and project-related data available in their profiles and software repositories. However, unlike in OSNs, the privacy concerns and practices of SCP users have not been extensively explored nor documented in the current literature. In this work, we present the preliminary results of an online survey (N=105) addressing developers' concerns and perceptions about privacy threats steaming from SCPs. Our results suggest that, although users express concern about social and organisational privacy threats, they often feel safe sharing personal and project-related information on these platforms. Moreover, attacks targeting the inference of sensitive attributes are considered more likely than those seeking to re-identify source-code contributors. Based on these findings, we propose a set of recommendations for future investigations addressing privacy and identity management in SCPs.
\end{abstract}

\begin{IEEEkeywords}
social coding platforms, privacy concerns, usable security, privacy engineering
\end{IEEEkeywords}

\section{Introduction} \label{sec:introduction}

Over the last years, the landscape of modern software engineering has significantly evolved thanks to the emergence of Social Coding Platforms (SCPs) like GitHub and BitBucket \cite{kinsman2021msr,mens2019social}. From software enthusiasts to more senior engineers, SCPs bring together individuals with diverse technical expertise and background beneath the premises of public and collaborative software development \cite{montandon2021ist}. Furthermore, since their introduction in the late 2000s, SCPs have not only attracted new players into the Open Source Software (OSS) ecosystem, but also redefined the cooperative nature of OSS deployment by offering free hosting and version control capabilities enhanced with social features \cite{cosentino_systematic_2017,meli_how_2019}.

\begin{figure}
\centering
\includegraphics[width=\linewidth]{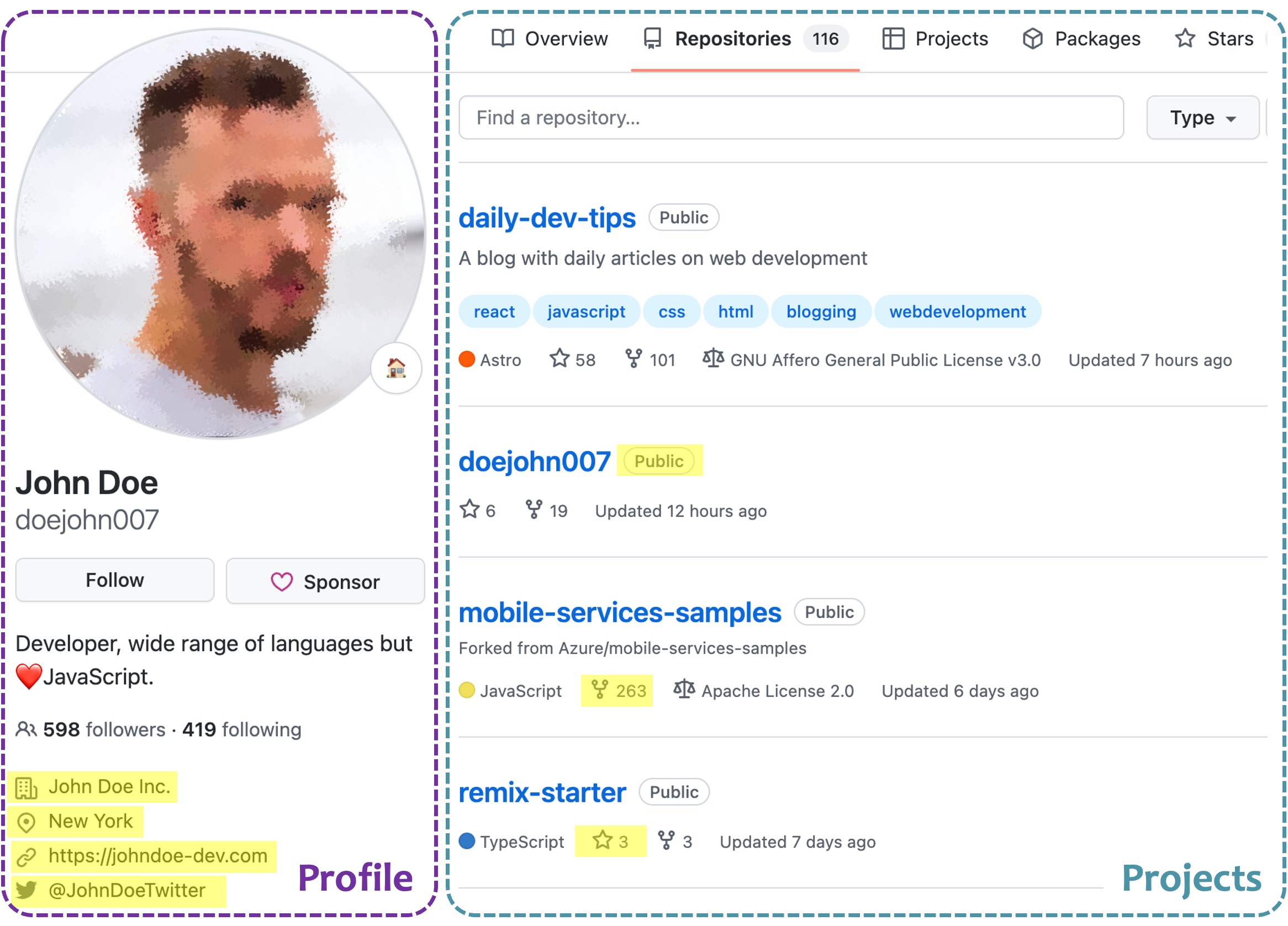}
    \caption{An example of a developer page on GitHub.}
    \label{fig:profile} \vspace{-2ex}
\end{figure}

At their core, SCPs resemble many characteristics of mainstream Online Social Networks (OSNs) like Facebook or Twitter. For instance, in order to network, collaborate, and disseminate their projects, SCP users often set up a profile like the one in Fig.~\ref{fig:profile}. Such profiles condense plenty of information about developers' demographics and their behaviour within the platform, including the projects and repositories they own, maintain, and contribute to. Moreover, some of the activity metrics one can obtain from a particular user include (i) the number of repositories they created, (ii) the number of issues they reported, (iii) the number of pull requests they submitted, and (iv) the number of commits they authored \cite{wu_exploring_2019}. 

Developer-centred SCP metrics (and the information they aggregate) have helped researchers to characterise the GitHub community and unveil common software engineering practices in the wild. Still, such information can also be leveraged to carry on targeted cyberattacks on specific users and socio-technical groups. That is, on individuals of a certain gender, region, or with a particularly technical background (e.g., users prone towards insecure coding practices) \cite{lazarine_identifying_2020}. Furthermore, with the current advances in Artificial Intelligence (AI) and Machine Learning (ML), even those keeping a more private, less visible profile could be re-identified thanks to the large amounts of data available on the platform.

\paragraph*{\textbf{Motivation}}

Privacy and identity management have been extensively investigated within the context of OSNs. Overall, the unintended disclosure of personal information, anonymity, and data misuse are some of the many issues addressed within the current literature \cite{oukemeni_privacy_2019}. Likewise, the inference of sensitive attributes from metadata and the re-identification of data subjects have received great attention among privacy researchers seeking to elaborate Privacy Enhancing Technologies (PETs) for online networking \cite{henriksen-reidentification_2016, beigi_survey_2020}. Particularly, for shaping socio-technical solutions aiming to protect users' identity and personal attributes from re-identification, inference, and linkability attacks.

Despite the similarities between SCPs and OSNs, the privacy issues of the former have not received as much attention as the ones of the latter. However, information available in SCP platforms can be leveraged to perform social engineering attacks and, in turn, compromise the security of companies and public agencies at large. For instance, in early 2022, a group of cyberattackers gained access to 190 Gb of Samsung's data using secret keys exposed in public repositories \cite{secweek2022}. Moreover, a recent report by GitGuardian \cite{gguardian2022} shows that the extent of publicly-exposed secrets on GitHub has more than doubled since 2020. Though alarming, such numbers should not be surprising as it just takes a developer inadvertently pushing code to a public repository to compromise the security of a whole system \cite{infosec2022}. Hence, there is a call for further investigations seeking to characterise the privacy threats and concerns emerging within collaborative coding environments.

\paragraph*{\textbf{Contribution and Research Questions}}

In this work, we present and discuss the preliminary results of an empirical study about privacy concerns in SCPs. Such a study consisted of an online survey (N=105) addressing the following Research Questions (RQs):
\begin{itemize}
    \item \textbf{RQ1}: \textit{How concerned are users of SCPs about their privacy?} To answer this RQ, we designed a questionnaire based on privacy-related constructs available in the current literature. Particularly, we assessed developers' perceived control over their personal and project-related information, their concerns about social/organisational threats, and their prior experiences of privacy invasion in SCPs, among others.
    \item \textbf{RQ2}: \textit{What is their perceived likelihood of certain privacy threats emerging within SCPs?} The purpose of this RQ is to determine whether SCP users underestimate the chances of experiencing privacy attacks. For this, participants were asked to assess the likelihood of two hypothetical privacy attack scenarios describing (i) the inference of sensitive attributes based on profile data and (ii) the re-identification of code authorship from information available in SCP repositories. %How likely do they perceive certain privacy threats emerging within SCPs?
\end{itemize}

All in all, our findings suggest the presence of concerns related to social and organisational privacy threats among SCP users. Still, many participants reported (i) feeling safe publishing personal and project-related data in these environments and (ii) having a sense of control over such data. We discuss the implications of these insights and elaborate on some perspectives for further research endeavours.

\section{Background and Related Work}

\subsection{Privacy Research in Online Social Networks}

Many investigations have addressed users' concerns and attitudes towards service providers, information access, and personal data sharing in OSNs. These efforts have yielded significant findings about the role such concerns play on people's self-disclosure behaviour, access-control preferences, and overall perception of online privacy risks \cite{oukemeni_privacy_2019}. For instance, it has been repeatedly shown that users of OSN often fail to preserve the contextual integrity of their personal data despite their high privacy concerns \cite{schwartz2020selectivity}. This phenomenon is usually associated with a low usability of security controls and the lack of transparency with which platforms collect, process and store users' data \cite{oukemeni_privacy_2019}. Recent publications have also drawn attention to the cognitive biases affecting people's privacy decision-making processes \cite{ostendorf2022theoretical}. Particularly, they suggest that users of OSNs are prone to replace rational risk judgements with ``rules of thumb'' or heuristics to reduce the complexity of their cybersecurity choices \cite{diaz2022enagram}. In turn, they often adopt sub-optimal privacy configurations and expose their data to untrusted audiences or malicious users.

Prior work has also sought to characterise potential privacy attacks in OSNs. Such attacks typically use information that users disclose in their profiles and posts to feed ML models capable of inferring sensitive attributes and behavioural patterns \cite{liu_when_2021}. For instance, Gu et al. \cite{gu_we_2016} developed a probabilistic model that employs user-centric and social-relationship data to infer individuals' home location. Such a model can achieve more than 90\% accuracy if the targeted user had disclosed some location data in the past (e.g., check-in at a bar or a restaurant) and around 60\% when not. Likewise, Comito \cite{comito_next_2020} introduced a supervised learning framework for predicting the next location of an individual based on her prior mobility behaviour. Recent publications also give an account of the effectiveness of Deep Neural Networks (DNNs) for detecting sensitive information on images, including peoples' age \cite{garain_gra_net_2021}, gender \cite{garain_gra_net_2021}, and race \cite{vo_race_2018}. Overall, DNN-based privacy attacks outperform other state-of-the-art approaches due to their high performance and efficiency over large datasets \cite{liu_when_2021}.

\subsection{Privacy Research in Social Coding Platforms} \label{sec:scp_research}
%\subsection{Privacy Threats in Social Coding Platforms}

Unlike in OSNs, research addressing SCPs has not placed much attention on users' concerns about privacy and data protection. Moreover, at its best, current investigations can be seen as attack models aiming to unveil correlations between developers' demographic attributes and behavioural patterns emerging within these platforms. For instance, Wu et al. \cite{wu_exploring_2019} observed that the number of pull requests and commits generated by GitHub users are linked to specific categories of profile pictures (e.g., human, logo, or animal) and demographic characteristics. Research addressing gender biases has also contributed to characterising developers' behaviour in SCPs. A study by Imtiaz et al. \cite{imtiaz_investigating_2019} showed that female users in GitHub are more restrained in communication than men and tend to concentrate their work across fewer projects and organisations. Likewise, Newton et al. \cite{newton_modeling_2022} found that GitHub contributors identified as women have a shorter tenure in open source projects compared to men. 

Research addressing code attribution has also helped, to a large extent, depict the landscape of privacy attacks in SCPs. Particularly, it has been shown that code available across public software repositories can be exploited to re-identify its contributors through stylistic patterns reflecting individual programming styles. These patterns range from simple artefacts in comments and code layouts to subtle habits in syntax usage and control flow. For instance, \citet{caliskan-islam_-anonymizing_2015} applied ML methods to de-anonymize the authors of C/C++  source code using a stylometry feature set extracted from Abstract Syntax Trees (ASTs). Alongside, \citet{alsulami_source_2017} introduced an LSTM model for the automatic extraction of such features, whereas \citet{abuhamad_large-scale_2018} proposed a similar technique targeting large-scale, language-oblivious, and obfuscation-resilient code authorship identification. This last one is a language-sensitive approach capable of identifying authors across four programming languages (e.g., C, C++, Java, and Python), even when using a combination of them (e.g., Java/C++, Python/C++). 

\section{Methodology}

\subsection{Survey Design}

We conducted an online survey to answer the research questions presented in Section~\ref{sec:introduction}. Such a survey had two main sections, the first addressing privacy concerns in SCPs (RQ1) and the second focusing on privacy attacks stemming from information available in profiles and code repositories (RQ2). For the elicitation of privacy concerns, we referred to the following psychological constructs and scales available in the current literature on OSNs:
\begin{itemize}
    \item \textbf{Perceived Control (CTRL)} over the personal and project-related information disclosed in SCPs \cite{krasnova2010online}.
    \item \textbf{Perceived Privacy Risks (RSK)} in SCPs \cite{krasnova2010online}.
    \item \textbf{Prior Invasion of Internal Privacy (PIIP)} in SCPs, such as privacy violations experienced in the past \cite{morlok2016sharing}.
    \item \textbf{Privacy Concerns on Organisational Threats (PCOT)}, including the collection and secondary usage of data by service providers and third parties \cite{Krasnova2009}.
    \item \textbf{External Informational Privacy Concerns (EIPC)} on information disclosure practices that may affect others' privacy in SCPs (e.g., colleagues and collaborators) \cite{morlok2016sharing}.
    \item \textbf{Privacy Concerns on Social Threats (PCST)}, such as embarrassment and potentially malicious actions performed by other SCP users \cite{Krasnova2009}.
\end{itemize}

These scales have been elaborated and validated by their corresponding authors: CTRL and RSK by Krasnova et al. \cite{krasnova2010online}; EIPC and PIIP by Morlok \cite{morlok2016sharing}; PCOT and PCST by Krasnova et al. \cite{Krasnova2009}. We have slightly adapted them for the purpose of this study by replacing the term ``OSNs'' with ``SCPs'' on each scale item. All constructs were measured using a 6-point Likert scale ranging from \textit{completely disagree} to \textit{completely agree}. We have assessed their reliability by computing the corresponding Cronbach's Alpha coefficients, which resulted higher than 0.7 in all cases. This suggests that the items of each scale have relatively high internal consistency since values above 0.7 are considered ``acceptable''.

The second part of the survey consisted of two hypothetical privacy threat scenarios that participants had to assess. As shown in Fig.~\ref{fig:scenarios}, such scenarios represent privacy attacks documented in the current literature that can be performed with information available in SCPs (Section~\ref{sec:scp_research}). That is, (i) finding sensitive correlations between developers' activities and their demographic characteristics and (ii) re-identify their source code contributions through stylistic coding patterns. Particularly, we asked participants how likely they think these scenarios could occur in real life using a scale with the values \textit{very likely}, \textit{likely}, \textit{somehow likely}, \textit{somehow unlikely}, \textit{unlikely}, and \textit{very unlikely}.

\begin{figure} \label{scenarios}
\begin{GrayBox}\small
\textbf{SCENARIO 1}: \textit{Your activity in SCPs (e.g., commits, comments, and reactions) can be used to infer your gender and other socio-demographic information.}\vspace{2ex}

\textbf{SCENARIO 2}: \textit{Someone (e.g., a colleague) pushes code you wrote into a public repository using his/her account. Then the authorship of this code can be traced back to you using information available in SCPs (e.g., other code snippets you wrote and your profile information).}
\end{GrayBox}\vspace{-1ex}
\caption{Hypothetical privacy attack scenarios in a SCP.}
\vspace{-2ex}
\end{figure}

\subsection{Population and Sampling} \label{sampling}
Participants were recruited via Prolific\footnote{\url{https://prolific.co/}} and pre-screened based on their (i) self-reported knowledge of software development techniques and (ii) self-reported computer programming skills. They had to be at least 18 years old to join the study and were rewarded with GBP 2.00 (payed trough Prolific) for a completed survey of 15 minutes duration on average. As a standard quality control, we targeted users who already took part in at least 10 other studies and had a minimum approval rate of 98\% \cite{salminen2021suggestions}. Two attention questions were also included in the survey to identify and discard answers from unengaged participants. As a result, we collected 105 valid responses from 110 full submissions. 

\subsection{Ethical Considerations} 

The study was conducted in accordance with the Declaration of Helsinki and approved by the Ethics Committee of German Association for Experimental Economic Research. All participants received information about the study procedure (including data privacy statements) and were asked to give their informed consent before joining the experiment. They were also given the possibility to withdraw at any time without their answers being recorded. Survey instruments, consent forms, and study results are available in the paper's \textbf{Replication Package}\footnote{\url{https://doi.org/10.5281/zenodo.7692654}}.

\begin{table*}[h]
\renewcommand{\arraystretch}{1.1}
\caption{Survey items excerpt.}
\label{tab:survey_items}
\footnotesize \centering
\resizebox{0.85\linewidth}{!}{
\begin{tabular}{ C{1.5cm} | C{9cm} | C{0.4cm} C{0.4cm} C{0.4cm} C{0.4cm} C{0.4cm} C{0.4cm}}
\toprule
\textbf{Construct} & \textbf{Analysed construct item} & \textbf{1} & \textbf{2} & \textbf{3} & \textbf{4} & \textbf{5} & \textbf{6} \tabularnewline \midrule
\textbf{CTRL} & {I feel in control over the personal and technical information\\ I provide on SCPs} & \gradient{0} & \gradient{9} & \gradient{8} & \gradient{33} & \gradient{43} & \gradient{12} \tabularnewline
\textbf{RSK} & {I feel safe publishing my personal information and project-related information on SCPs} & \gradient{4} & \gradient{5} & \gradient{22} & \gradient{37} & \gradient{29} & \gradient{8} \tabularnewline
\textbf{PIIP} & I have already experienced\\ a violation of my privacy on SCPs & \gradient{44} & \gradient{42} & \gradient{9} & \gradient{5} & \gradient{4} & \gradient{1}
\tabularnewline
\textbf{PCOT} & {I am often concerned other parties could actually collect my publicly available information on SCPs} & \gradient{6} & \gradient{20} & \gradient{20} & \gradient{32} & \gradient{21} & \gradient{6} \tabularnewline
\textbf{EIPC} & {I am concerned that unauthorised people may access the code of my colleagues I shared on SCPs} & \gradient{9} & \gradient{17} & \gradient{23} & \gradient{37} & \gradient{14} & \gradient{5} \tabularnewline
\textbf{PCST} & {I am concerned that other users may take advantage of me based on the information they learned about me through SCPs} & \gradient{14} & \gradient{24} & \gradient{23} & \gradient{26} & \gradient{16} & \gradient{2} \tabularnewline
\bottomrule
\multicolumn{8}{l}{\rule{0pt}{2ex}\underline{Note}: 1= \textit{Completely disagree}, 2= \textit{Disagree}, 3= \textit{\textit{Somehow disagree}}, 4= \textit{Somehow agree}, 5= \textit{Agree}, 6= \textit{Completely agree}.}
\end{tabular}}
\vspace{-2ex}
\end{table*}

\begin{figure}
\centering\includegraphics[width=\linewidth]{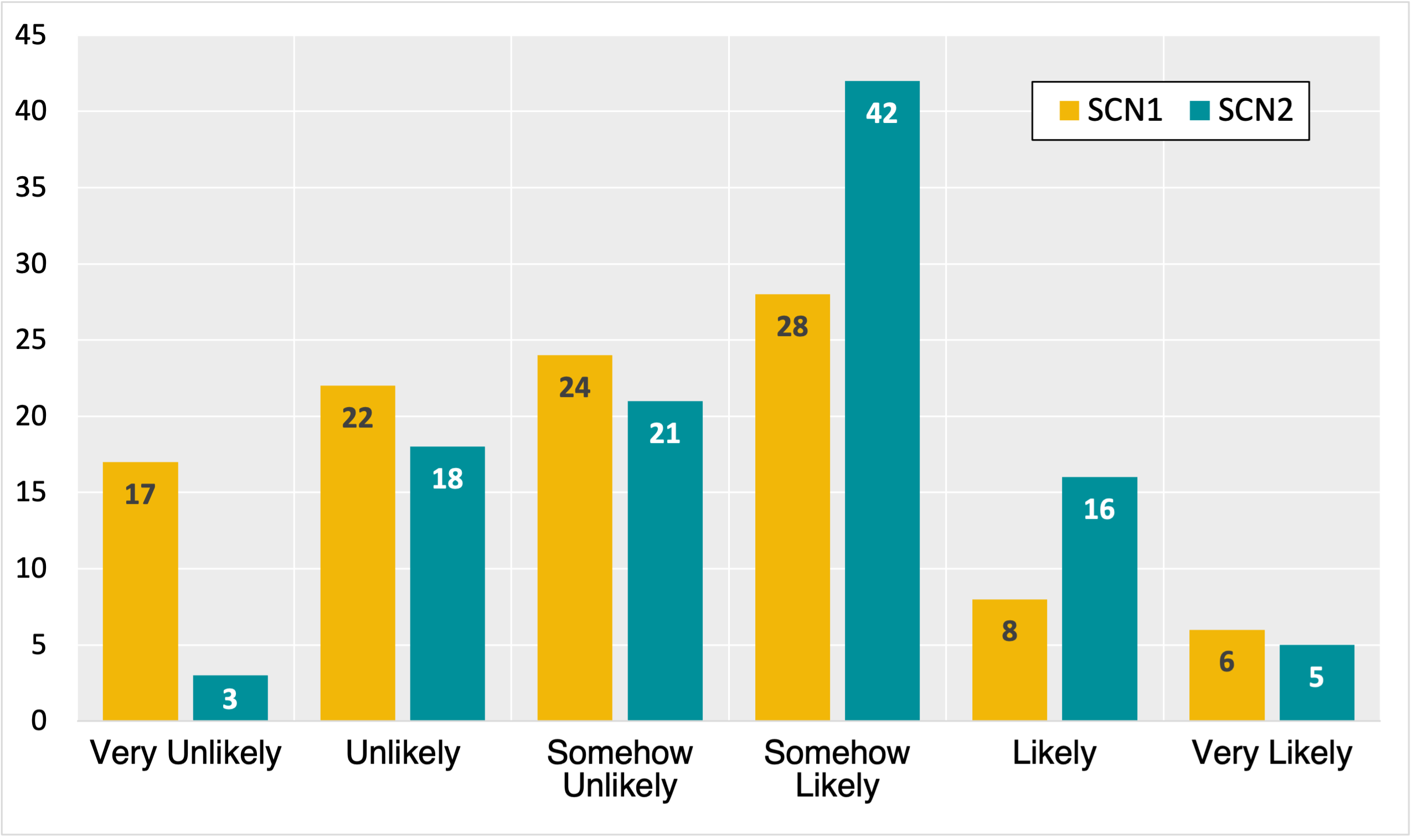}
    \caption{Assessment of privacy attack scenarios.} \vspace{-2ex}
    \label{fig:scenarios}
\end{figure}

\section{Results and Discussion}

Of 105 study participants, 94 were male, 10 were female, and 1 was non-binary. Around 48.6\% reported having 2 to 5 years of experience using SCPs, 14.3\% between 6 to 10, 13.3\% less than 2, and 3.8\% more than 10 years. Regarding software development, 35.2\% had less than 2 years of experience, another 35.2\% between 2 and 5, 16.2\% between 6 and 10, and 13.3\% more than 10 years. A complete overview of the sample's demographics is shown in Table~\ref{tab:demographics}.

Each of the constructs elicited in the first part of the survey (i.e., CTRL, RSK, PIIP, PCOT, EIPC, and PCST) contains between 3 and 7 statements or items. We have selected one representative item per construct to conduct a preliminary analysis of developers' privacy concerns in SCPs. Though exploratory, such an approach meets the purpose of this work, which is to provide some initial insights and motivate further investigations in this regard. A complete overview of the results obtained for each scale item is available in the paper's Replication Package.

As shown in Table~\ref{tab:survey_items}, most participants agreed with the CTRL statement about feeling in control over the personal and technical information they disclose in SCPs. Likewise, about 71\% agreed with the corresponding RSK statement (scores 4-6), whereas 29\% reported not feeling safe publishing personal and project-related information (scores 1-3). Regarding prior experiences, only 9.5\% of respondents had suffered some privacy violation in the past, while the rest tended to disagree with this construct item. On the other hand, about 59\% of the study participants reported concerns related to data collection from third parties (PCOT), and 56\% were worried about unauthorised access to others' code they had shared (EIPC). Finally, when it came to social threats, 42\% tended to agree with the PCST statement referring to the exploitation of publicly-available information by other SCP users.

Fig.~\ref{fig:scenarios} illustrates participants' assessment of the proposed privacy threat scenarios. We observe that 60\% placed Scenario~1 (SCN1) in the ``unlikely'' side of the scale (i.e., \textit{very unlikely}, \textit{unlikely}, or \textit{somehow unlikely}), whereas 40\% deemed it ``likely'' to a certain degree (i.e., \textit{somehow likely}, \textit{likely}, or \textit{very likely}). Surprisingly, these values get reversed when it comes to Scenario~2 (SCN2), where 60\% of participants considered it to some extent ``likely'' and the remaining 40\% ``unlikely''. When analysing each likelihood value separately, we can also spot some considerable differences between SCN1 and SCN2. Particularly, a larger number of participants considered SCN1 to be \textit{very unlikely} as opposed to SCN2, though fewer times \textit{likely} and \textit{somehow likely}.

\begin{table}[t]
%\section{Studied Sample} \label{sample}
\caption{Survey Self-Reported Demographic Data.}
\label{tab:demographics}
\footnotesize \centering
\begin{tabularx}{\linewidth}{lXcc}
\toprule
\textbf{Demographic} & \textbf{Ranges} & \textbf{Freq.} &\textbf{\%} \\
\midrule
\multirow{3}{*}{Gender}
& Male          & 94    & 89.5\%   \\
& Female        & 10    & 9.5\%   \\
& Non-Binary    & 1     & 1.0\%   \\          
\midrule
\multirow{4}{*}{\makecell[l]{Educational\\level}}
& High School or Less           & 14    & 13.3\%   \\
& Some College                  & 19    & 18.1\%   \\
& Undergraduate (BSc,~BA)       & 48    & 45.7\%   \\
& Graduate (MSc, PhD)           & 24    & 22.9\%   \\
\midrule
\multirow{6}{*}{\makecell[l]{Employment\\status}}
& Student                               & 26    & 24.8\%   \\
& Unemployed, looking for work       & 10    & 9.5\%   \\
& Unemployed, not looking for work   & 2     & 1.9\%   \\ 
& Working full-time                     & 13    & 12.4\%   \\ 
& Working part-time                     & 54    & 51.4\%   \\        
\midrule
\multirow{4}{*}{\makecell[l]{Software\\development\\experience}}
& $<$2 years        & 37    & 35.2\%   \\
& 2-5 years         & 37    & 35.2\%   \\ 
& 6-10 years        & 17    & 16.2\%   \\
& $>$10 years       & 14    & 13.3\%   \\               
\midrule                
\multirow{4}{*}{\makecell[l]{Experience\\using SCPs}}
& $<$2 years        & 35    & 13.3\%   \\
& 2-5 years         & 51    & 48.6\%   \\ 
& 6-10 years        & 15    & 14.3\%   \\
& $>$10 years       & 4     & 3.8\%   \\  
\bottomrule
\end{tabularx}
\vspace{-2ex}
\end{table}

\subsection{Implications for Privacy Research}

\textbf{(i) Privacy Decision-Making}: Our results suggest that, although users of SCPs usually feel in control of their information, they are also concerned about privacy threats stemming from (i) data collection by service providers and third parties, (ii) unauthorised access to code authored by other users, and (iii) the exploitation of publicly-available information. Prior investigations in OSNs have also reported a similar offset phenomenon often referred to as the ``privacy paradox'' \cite{barth2019putting}. That is, a discrepancy between users' privacy concerns and actual privacy-related behaviour \cite{kim2020understanding}. Nevertheless, a feeling of control normally contributes to mitigating these concerns and ultimately to higher levels of information disclosure \cite{kim2020understanding}. Hence, this observation should be further investigated and discussed with the help of additional empirical evidence. 

\textbf{(ii) Usability and Transparency}: Participants' answers to the PCOT, EIPC, and PCST items also raise questions about the usability of current privacy-enhancing mechanisms in SCPs. The perception of organisational threats is often a product of a lack of transparency about how (personal) information is collected, processed, stored, and shared by the platform \cite{wilkinson2021pursuit}. In this regard, it would be beneficial to analyse the extent to which SCP users have a fair understanding of common data-processing practices (e.g., from SCPs and third parties) and how their knowledge is translated into more (or less) restrictive privacy configurations (e.g., visibility of software repositories). To some degree, similar concerns have also been reported by practitioners within the Mining Software Repositories (MSR) community \cite{gold2022ethics, vidoni2022ropes}. Particularly on the ethical implications of retrieving information from SCP repositories without obtaining full consent from their users. It is thus critical not only to review the potentially invasive nature of these practices, but also to assess the limitations and drawbacks of current access-control mechanisms in SCPs.

\textbf{(iii) Multi-Party Privacy}: Users' social and external privacy concerns have motivated recent investigations addressing the resolution of multi-party privacy conflicts in OSNs \cite{humbert2019csur}. Such conflicts often arise when users share information that can also compromise the privacy of others, such as a group picture or post mentions. All in all, multi-party privacy takes a collective view on the norms and boundaries of information disclosure, putting special emphasis on the conflicting privacy preferences among the co-owners of particular data items \cite{such2018multiparty}. Secrets sprawl across software repositories can be considered as an instance of such conflicts in SCPs. Furthermore, pull request-based workflows could help (in principle) to prevent secret sharing as they promote the monitoring and reviewing of code being pushed to a particular repository \cite{krause22acc}. Still, specific sources of multi-party privacy conflicts, concerns and countermeasures should be systematically investigated. %in future studies.

\textbf{(iv) Risk Assessment}: Prior research has shown that past experiences can shape, to a great extent, people's privacy perception and behaviour \cite{renaud2022regrets}. Therefore, it is not surprising that a similar number of participants tended to disagree with the PIIP statement but also agree with the RSK statement (90\% and 70\%, respectively). Furthermore, participants' assessment of the proposed threat scenarios is also aligned with their overall perception of privacy control and safeness, as many believed SCN1 and SCN2 to be \textit{very unlikely}, \textit{unlikely}, or \textit{somehow unlikely}. Given the central role of risk perception in privacy decision-making \cite{ostendorf2022theoretical}, it is critical to raise awareness among developers about the cybersecurity threats emerging within SCPs. Equally important is to endow them with methods and tools that help preserve the contextual integrity of the information they share while assessing the potential~risks. %help them

\subsection{Limitations and Threats to Validity}

The present work is subjected to limitations related to the size and composition of the studied sample. On the one hand, its relatively small size does not allow us to generalise the study results to the whole SCP community. Hence, the insights gained from it should be seen as preliminary instead of conclusive and motivate further investigations in this area. On the other hand, participants recruited via crowd sourcing platforms may incur in dishonest practices such as falsely claiming to meet the experiment's eligibility criteria or provide random answers. As mentioned in Section~\ref{sampling}, we have sought to minimise these threats by employing Prolific's built-in qualification features to recruit suitable participants who may actively engage in our study. In the future we plan enhance this step by applying expertise questions like the ones proposed by \citet{danilova2021you}.

\section{Conclusion and Future Work}

SCPs are rich sources of information about developers' sociotechnical skills and have helped, to a great extent, characterise current trends in software engineering. However, despite their importance, little attention has been placed on the privacy-related behaviour of SCP users and the threats steaming from unsavvy information disclosure practices. In this work, we provided some empirical insights in this regard with the aim of paving the road for future investigations. Overall, our results suggest the need for further research assessing the usability of the current privacy-enhancing mechanism available in SCPs. That is, to determine whether such mechanisms meet users' individual and collective goals of transparency, access control, and anonymity.

Developers' knowledge and perception of privacy threats also call for additional research efforts, as it is critical to assess their (potential) lack of awareness in order to outline adequate cybersecurity training programs. The role of risk awareness in the adoption of privacy-enhancing technologies has been thoroughly studied and documented across the OSN literature \cite{oukemeni_privacy_2019}. In future work, we plan to delve into the interplay between risk awareness and privacy-related behaviour in SCPs. Particularly, on developers' knowledge about potential privacy threats (e.g., secrets sprawl or code re-attribution) and their adoption of cybersecurity best practices (e.g., see \cite{krause22acc}).

\newpage

\bibliographystyle{IEEEtranN}

{\footnotesize
\bibliography{references.bib}}

\end{document}